\documentclass[twocolumn,pra,showpacs,nofootinbib,superscriptaddress]{revtex4}

\usepackage{dcolumn,bm,graphicx,amsmath}
\usepackage
{hyperref}

\begin{document}
\preprint{UNR Jan 2005-\today }
\title{ Molecular CP-violating magnetic moment}

\author{Andrei Derevianko}
\affiliation{Department of Physics, University of Nevada, Reno,
Nevada 89557}

\author{M. G. Kozlov}
\affiliation{Petersburg Nuclear Physics Institute, Gatchina
188300, Russia}

\date{\today}

\begin{abstract}
A concept of  CP-violating (T,P-odd) permanent molecular magnetic moments
$\mu^\mathrm{CP}$ is introduced. We relate the moments to the electric dipole
moment  of electron (eEDM) and  estimate $\mu^\mathrm{CP}$ for several
diamagnetic polar molecules. The moments exhibit a steep, $Z^5$, scaling with
the nuclear charge $Z$ of the heavier molecular constituent. A measurement of
the CP-violating magnetization of a polarized sample of heavy molecules may
improve the present limit on eEDM by several orders of magnitude.
\end{abstract}


\maketitle

It is a common knowledge, that  heteronuclear diatomic molecules possess a
static electric dipole moment aligned with the internuclear axis
$\hat{\bm{n}}$, $\bm{D}=D \, \hat{\bm{n}}$. For a diamagnetic molecule,
however, there is no similar magnetic moment. As demonstrated below, an
existence of such a magnetic moment would violate both parity-transformation
(P) and time-reversal (T) discrete symmetries. Because of the compelling CPT
theorem, an observation of this magnetic moment would provide an evidence for
the CP violation~\cite{KhrLam97,BigSan00}. CP-violation, although observed in
particle physics, still remains a mystery, as much stronger CP-violating
mechanisms may be required to explain the matter-antimatter asymmetry of the
Universe.

Here we introduce the molecular CP-violating magnetic moments,
$\bm{\mu}^\mathrm{CP}=\mu^\mathrm{CP} \hat{\bm{n}}$. We propose a measurement
of $\mu^\mathrm{CP}$ via detection of ultra-weak magnetic fields generated by a
polarized sample of diamagnetic molecules. For several molecules  we evaluate
$\mu^\mathrm{CP}$ and express them in terms of the permanent electric dipole
moment (EDM) of electron, $d_e$. A measurement of non-vanishing molecular
CP-violating moments would reveal the elusive electron EDM (eEDM) (EDMs violate both T
and P symmetries). While no EDMs have been found so far, it is anticipated that
the next generation of experiments may finally discover the  EDMs. Indeed, most
supersymmetric extensions of the Standard Model of elementary particles predict
eEDMs that are within a reach of planned and on-going experimental
searches (see, e.g., a popular review~\cite{ForPatBar03}). The present limit on
$d_e$ comes from an atomic Tl beam experiment~\cite{RegComSch02},
\begin{equation}
d_e\left( \mathrm{Tl} \right) < 1.6 \times 10^{-27} e\cdot
\mathrm{cm}\, . \label{Eq:deTl}
\end{equation}
Here we propose an experimental search for the CP-violating magnetic moments of
heavy polar molecules. We argue that the limit on $\mu^\mathrm{CP}$ derived
from such experiments would imply constraints on $d_e$ that are several
orders of magnitude better than the present limit~(\ref{Eq:deTl}).
In principle, the experiments can be carried out with any diamagnetic polar molecules.
However, there is no particular advantage in using polyatomics,
and we restrict our consideration to polar diatomic molecules.

In the remainder of this paper, unless specified otherwise, we use atomic units
$|e|=\hbar=m e\equiv 1$ and the Gaussian system for electro-magnetic equations. In
these units, the Bohr magneton is $\mu B=\alpha/2$, where $\alpha\approx 1/137$
is the fine structure constant, and the unit of magnetic field is $m e^2
e^5/\hbar^4 \approx 1.72 \times 10^7 \, \mathrm{Gauss}$.

{\em General considerations.}
Diatomic molecule is characterized
by the projection $\Omega=(\bm{J} \cdot \hat{\bm{n}})$ of the total electronic
angular
momentum $\bm{J}=\bm{L}+\bm{S}$ on the internuclear axis $\hat{\bm{n}}$.
For a molecular state with a definite $\Omega$,
the molecular magnetic
moment is directed along $\hat{\bm{n}}$ and, phenomenologically, we may
construct the following combinations of the two vectors
\begin{equation}
\bm{\mu} = \mu^\mathrm{CP} \hat{\bm{n}} + \mu_B G_\parallel (\bm{J} \cdot
\hat{\bm{n}}) \hat{\bm{n}},
\label{Eq:muGeneral}
\end{equation}
where $\mu^\mathrm{CP}$ and $G_\parallel$ are numbers. For the  Hund's case (a)
$G$-factor is given by an expression $G_\parallel\Omega\approx\Lambda+2\Sigma$,
where $\Lambda=(\bm{L} \cdot \hat{\bm{n}})$ and $\Sigma=(\bm{S} \cdot
\hat{\bm{n}})$~\cite{LanLif97}. While the second term in (\ref{Eq:muGeneral})
is T,P-even, the $\mu^\mathrm{CP} \hat{\bm{n}}$ term violates both
time-reversal (and thus CP) and parity. Indeed, under the time reversal the
magnetic moment acquires a minus sign, while $\hat{\bm{n}}$ is T-invariant.
Similarly, under parity transformation, $\bm{\mu}$ is not affected, while
$\hat{\bm{n}}$ flips direction.


Given a complete set of molecular states $|k\rangle$ (with energies $E_k$),
the magnetic moment
$\mu^{\mathrm{CP}}$ of
a state $|0\rangle$ can be
computed as
\begin{equation}
\mu^{\mathrm{CP}}=2\sum_k \frac{\langle 0|\left(
\bm{M}\cdot\hat{\bm{n}}\right) |k\rangle\langle k|V^{\mathrm{CP}}| 0
\rangle}{E_0-E_k} \, , \label{Eq:muSum}
\end{equation}
where $\bm{M}$ is the operator of magnetic dipole moment, and the
CP-violation is due to a T,P-odd interaction $V^\mathrm{CP}$. Here
we consider eEDM as a source of  CP-violation, so that
$V^{\mathrm{CP}}=-d_e(\gamma_0-1)\gamma_0\gamma_5 \bm{\gamma} \cdot
\bm{\mathcal{E}}_\mathrm{int}$, where $\gamma_i$ are the
conventional Dirac matrices~\cite{KhrLam97}. The electric field
$\mathcal{E}_\mathrm{int}$ exerted upon the EDM is the strongest at
the nucleus, leading to $\bm{\mathcal{E}}_\mathrm{int} \approx
(Z/r^2) \, \hat{\bm{r}}$, where $Z$ is the nuclear charge and
$\bm{r}$ is the radius-vector of the electron with respect to the
nucleus. The matrix element of $V^\mathrm{CP}$ scales as $Z^3$, so
that the interaction $V^\mathrm{CP}$ can be considered as localized
at the heavier nucleus.
Below we will evaluate the molecular sum~(\ref{Eq:muSum}) using an
approach similar to the LCAO method (linear combination of atomic
orbitals).

Note that Eq.~(\ref{Eq:muSum}) is expressed in the body-frame of the
molecule. After $\mu^{\mathrm{CP}}$ is found one has to average
Eq.~\eqref{Eq:muGeneral} over rotations. In the external electric
field $\langle\bm{n}\rangle \ne 0$ and we get magnetization in the
direction of the electric field. The second T,P-even term in
Eq.~\eqref{Eq:muGeneral} does not contribute to this magnetization.
For diamagnetic molecules $(\bm{J} \cdot \hat{\bm{n}})=\Omega=0$ and
this term vanishes. For paramagnetic molecules in the absence of the
magnetic the levels with different signs of $\Omega$ are equally
populated and this term is averaged to zero.

We would like to detect a magnetization of a sample of polarized molecules due
to CP-violating magnetic moments. In this regard, it is beneficial to work with
diamagnetic molecules $\Omega=0$, so that the traditional T,P-conserving
magnetic moment (the last term in Eq.~(\ref{Eq:muGeneral})) does not contribute
to the magnetization. Some of the molecules may still have non-zero {\em
nuclear} magnetic moments. However, the magnetization due to the nuclear
moments in a macroscopic sample will average out to zero, since the nuclear
spins have equal probabilities of orienting parallel or anti-parallel to the
direction of internuclear axis. Another advantage of diamagnetic molecules is
that they are also chemically stable allowing for higher sample densities and
thus for a larger sample magnetization.

To illustrate our qualitative approach to evaluating CP-violating magnetic
moments, consider a polar molecule CsF in its ground $^1\!\Sigma$ state.
Halides exhibit a chemical bond of a strong ionic character, and we model the
CsF  molecule as the Cs$^+$ ion perturbed by the electric field $\mathcal{E}$
of negative ion F$^-$. The perturbing field at the Cs$^+$ is $\mathcal{E}
\approx q/R_e^2$, where $R_e$ is the internuclear separation and $q=1$ is the
valency of Cs. The CP-violation is enhanced near the heavier atom and we may
evaluate the magnetic moment as
\begin{equation}
\mu^\mathrm{CP}(\mathrm{CsF}) \approx \beta^\mathrm{CP}(\mathrm{Cs}^+)
\frac{q}{R_e^2},
\label{Eq:MuEstimate}
\end{equation}
where $\beta^\mathrm{CP}(\mathrm{Cs}^+)$ is a so-called CP-violating
polarizability~\cite{Bar04,RavKozDer05} of the Cs$^+$ ion. Thus the molecular
two-center problem is reduced to computing a one-center property ---
CP-violating polarizability of the heavier constituent. If both constituents of
the diatomic molecule AB have comparable nuclear charges, then
$\mu^\mathrm{CP}(\mathrm{AB}) \approx [\beta^\mathrm{CP}(\mathrm{A}^{(+q)})
 - \beta^\mathrm{CP}(\mathrm{B}^{(-q)})]\,q R_e^{-2}$, where $q$ is the observed
valency of the atoms.

{\em CP-violating polarizabilities.} The effect of CP-violation on
electromagnetic properties of the media has been considered by a number
of authors~(see, e.g., \cite{Sha68,Mos86,BiaModWil86,Bar04}). In
particular, the atomic CP-violating (T,P-odd) polarizability
$\beta^\mathrm{CP}$ relates induced atomic magnetic moment
$\mu^\mathrm{CP} \mathrm{at}$ to the externally applied electric field
$\mathcal{E}$,
\begin{equation}
\label{Eq:BetaCPDef}
\mu^\mathrm{CP}_\mathrm{at}=\beta^\mathrm{CP} \mathcal{E}.
\end{equation}
For a spherically-symmetric system,
the CP-odd polarizability is a scalar quantity, i.e.,
the induced magnetic moment is directed along the applied E-field.

In Ref.~\cite{RavKozDer05} we have computed and explored the atomic
CP-violating polarizabilities for rare-gas atoms in their respective $^1\!S_0$
ground states. Since the calculation reported here builds upon that work, let
us briefly recapitulate our approach and main results. We related
$\beta^\mathrm{CP}$ to the eEDM  through third-order perturbation
theory: the CP-odd polarizability of an atomic state $\Phi_0$ can be
represented as
\begin{eqnarray}
  {\beta}^\mathrm{CP} &=&
-2 \langle \Phi_0 |V^\mathrm{CP}\,\mathcal{R}\,
M_{z}\,\mathcal{R}\,D_z|\Phi_0\rangle \nonumber
\\
&&- 2\langle
\Phi_0|M_{z}\,\mathcal{R}\,V^\mathrm{CP}\mathcal{R}\,D_z|\Phi_0\rangle
\label{Eq:MuResolvent}
\\
&&-2\langle
\Phi_0|M_{z}\,\mathcal{R}\,D_z\,\mathcal{R}\,V^\mathrm{CP}|\Phi_0\rangle \, ,
\nonumber
\end{eqnarray}
where the resolvent operator
$\mathcal{R}= \left(\varepsilon_{0}-H_\mathrm{at}\right)^{-1}$,
$\varepsilon_{0}$ being the energy of the state $\Phi_0$ and $H_\mathrm{at}$
being the atomic Hamiltonian. We have evaluated Eq.(\ref{Eq:MuResolvent}) using
the Dirac-Hartree-Fock (DHF) approximation. Numerical evaluation has been carried
out  using a B-spline basis set technique\cite{JohBluSap88}.

In Ref.~\cite{RavKozDer05}, we demonstrated that  $\beta^\mathrm{CP}$ has a
doubly relativistic origin: relativistic effects are essential for
non-vanishing matrix elements of $V^\mathrm{CP}$ (Schiff theorem) and also due
to peculiar properties of magnetic-dipole operator $M$. This doubly
relativistic nature leads to a steep, $Z^5 R(Z)$, scaling of
$\beta^\mathrm{CP}$ with the nuclear charge, where slowly-varying $R(Z)$ is a
relativistic enhancement factor (see Ref.~\cite{RavKozDer05} for details). This
result, together with the estimate~(\ref{Eq:MuEstimate}), immediately provides
us with the $Z$-scaling of the CP-violating magnetic moments of diamagnetic
molecules
\[
\mu^\mathrm{CP} \propto Z^5 R(Z) \, ,
\]
where $Z$ is the nuclear charge of the heavier constituent. This $Z$-scaling is
more substantial than the usual $Z^3 R(Z)$ scaling~\cite{San65} of the effects
of eEDM on the energy levels of paramagnetic molecules (radicals). As
in the traditional EDM searches, it is beneficial to search for non-vanishing
CP-odd magnetic moments with molecules involving heavy atoms.
Diamagnetic contribution \eqref{Eq:MuResolvent} to CP-odd polarizability is
universal. However, for the systems with unpaired electrons this
contribution is masked by a larger paramagnetic term which is linked to
the electronic angular momentum \cite{Sha68}. The
advantage of the diamagnetic systems is the much lower magnetic noise,
which may be crucial for the experiment. On the theoretical side,
calculation of $\mu^\mathrm{CP}$ for diamagnetic molecules may be
simpler and more reliable than for para- or ferromagnetic crystals.

{\em Results of calculations}. In Table~\ref{table:betaValues} we present the
results of our calculations of CP-violating magnetic moments for several
diatomics: CsF, BaO,  TlF, PbO, and BiF. These diamagnetic molecules possess
the $^1\!\Sigma$ ground state. The heavier atoms of these diatomic pairs are
metals, and we assume that the molecules exhibit a pure case of ionic bond,
i.e., these heavier atoms fully lend their valence electrons to their
electronegative companions (F and O) and become closed-shelled $^1\!S 0$ ions.
The second and third columns of the Table list the resulting heavy atomic ions
with their nuclear charges, and in the fourth column we present our computed
values of CP-violating polarizabilities of these ions. Finally, we combine
ionic $\beta^\mathrm{CP}$ with the equilibrium internuclear separations (see
Eq.(\ref{Eq:MuEstimate})) and obtain an estimate for the molecular CP-violating
magnetic moments. Our sign convention in expression
$\bm{\mu}^\mathrm{CP}=\mu^\mathrm{CP} \hat{\bm{n}}$ is such that the unit
vector $\hat{\bm{n}}$ is directed from the heavier to the lighter nucleus.
Notice that we express the $\mu^\mathrm{CP}$ in terms of eEDM.
Similarly, $\mu^\mathrm{CP}$ can be expressed in terms of the T,P-odd
electron-nucleon couplings; a simple prescription of how to relate the present
results to the strength of such couplings is  given in Ref.~\cite{RavKozDer05}.

\begin{center}
\begin{table}[ht]
\begin{tabular}{llcdd}
\hline\hline  \multicolumn{1}{c}{Molecule} & \multicolumn{1}{l}{Ion}
& \multicolumn{1}{c}{$Z \mathrm{Ion}$}
& \multicolumn{1}{c}{$\beta^\mathrm{CP}(\mathrm{Ion})/d_e$}
& \multicolumn{1}{c}{$\mu^\mathrm{CP}/d_e$}    \\
\hline
CsF & Cs$^{+}$  & 55  & -3.0[-2]  &  1.5[-3]\\
BaO & Ba$^{++}$ & 56  & -2.3[-2]  &  3.4[-3]\\
TlF & Tl$^{+}$  & 81  &  2.9[-1] & -1.9[-2]\\
PbO & Pb$^{++}$ & 82  &  3.2[-1] & -4.9[-2]\\
BiF & Bi$^{+}$  & 83  &  4.8     & -3.2[-1]\\
\hline\hline
\end{tabular}
\caption{ Molecular CP-violating magnetic moments, $\mu^\mathrm{CP}/d_e$,
divided by the eEDM for several diamagnetic molecules. The values of
$\mu^\mathrm{CP}/d_e$ are dimensionless, $d_e$ and $\mu^\mathrm{CP}$  being
expressed in the Gaussian atomic units. The second, third, and the fourth
columns  list the heavier ion in the molecule, its nuclear charge and its
CP-violating polarizability, $\beta^\mathrm{CP}/d_e$, in units of
$\hbar^4/(m e^2 e^5)$. Notation $x[y]$ stands for $x \times 10^y$.
\label{table:betaValues} }
\end{table}
\end{center}

For all the considered molecules, the internuclear separation $R_e \approx
2$\AA, and thus the internal molecular fields are comparable.
More significant is the effect of increasing CP-violating
polarizabilities (the fourth column of Table~\ref{table:betaValues}) as one
progresses to heavier elements. This trend is largely due to the $Z^5$ scaling
of $\beta^\mathrm{CP}$. Yet, there is an order of magnitude of difference
between $\beta^\mathrm{CP}$ for Pb$^{++}$ ($Z=82$) and Bi$^{+}$ ($Z=83$). A
part of this large enhancement lies in a softer excitation spectrum of Bi$^{+}$
and thus smaller energy denominators in Eq.~(\ref{Eq:MuResolvent}). Also, while
solving the DHF equations we assumed that the outer shell of Bi$^{+}$ ion has
the $6p_{1/2}^2$ electronic configuration. In general, however, the ground
state of Bi$^{+}$ would contain a combination of $6p_{1/2}^2$ and $6p_{3/2}^2$
configurations. Since the $p_{1/2}$ states couple to EDM strongly, while
$p_{3/2}$ orbitals contribute at a much smaller level, we expect that our
result for $\beta^\mathrm{CP}$  of Bi$^{+}$ is somewhat overestimated.

Results of Table~\ref{table:betaValues} should be considered as a qualitative
estimate for another reason as well. The expressions~(\ref{Eq:MuEstimate}) and
(\ref{Eq:MuResolvent}) are based on atomic wavefunctions $|\Phi_i\rangle$,
instead of the molecular wavefunctions of the defining
expression~(\ref{Eq:muSum}). An underlying assumption is that the molecular
wavefunctions $|i\rangle$ in the vicinity of the heavier atomic ion can be
expressed perturbatively as
\[
|i\rangle \approx \left( |\Phi_i\rangle +
\sum_k \frac{\langle \Phi_k|-D_z \mathcal{E}| \Phi_i \rangle }
{\varepsilon_i -\varepsilon_k} \,
| \Phi_k \rangle  \right) |\Psi_0\rangle \, ,
\]
where $|\Psi_0\rangle$ is the wavefunction of the lighter ion (we left out
excitations from $|\Psi_0\rangle$ as being non-essential for computing
$\mu^\mathrm{CP}$). Certainly, {\em ab initio} molecular-structure calculations
of CP-violating magnetic moments are desirable to depart from this simple
perturbative picture. To motivate more sophisticated calculations, let us
evaluate a feasibility of an experimental determination of $\mu^\mathrm{CP}$
based on our qualitative estimates.

{\em Proposed experiments.} The molecular CP-odd magnetic moments are tiny. For
example, combining the present limit on the eEDM~(\ref{Eq:deTl}) with
the computed value of $\mu^\mathrm{CP}/d_e$, we obtain for BiF,
\begin{equation}
 \mu^{\mathrm{CP}} (\mathrm{BiF}) < 2.4 \times 10^{-37} \mathrm{erg/Gauss} \, .
 \label{Eq:muCPBiFpresent}
\end{equation}
While this is a remarkably small value, only $2.6 \times 10^{-17}$ of the electron
magnetic moment, measuring
such small magnetic moments seems to be possible with
the modern magnetometry.

The CP-violating magnetic moments can be determined by measuring the ultraweak
magnetic field generated by a sample of molecules. Because of rotations, the
body-fixed $\bm{\mu}^\mathrm{CP}$ moment  averages to zero in the laboratory
frame. Experimentally, one needs to apply a polarizing electric field
$\mathcal{E}_\mathrm{pol}$ to orient the molecules along the field. For the
efficient polarization, the coupling to the field must be stronger than the
rotational spacing, $D \mathcal{E}_\mathrm{pol} > 2 B$, where $B$ is the
rotational constant. For the ground vibrational state of  BiF, the rotational
constant is $B \approx 0.231\, \mathrm{cm}^{-1}$, requiring the application of
the polarizing E-field of a few kV/cm. Another consideration comes from thermal
averaging over rotational levels~\cite{VarGorEzh82}, and it is beneficial to
work at very low temperatures to have the smallest possible number of populated
rotational levels.

Suppose we have a spherical cell with polarized diamagnetic molecules; the permanent
CP-violating magnetic moments of the molecules  produce a macroscopic magnetization
of the sample generating an ultraweak magnetic field.
The maximum value of the magnetic
field at the surface of the cell can be expressed as
\begin{equation}
\mathcal{B}_{\max}= \frac{8\pi}{3} n \, \mu^\mathrm{CP} \, ,
\label{Eq:Signal}
\end{equation}
where $n$ is the number density of the sample. One could measure this induced
magnetic field and set the limits on the eEDM or other CP-violating
mechanisms. Clearly, one should increase the number density to enhance the
signal. However, condensing polar molecules with ionic bonds leads to a
crystallization of the sample. To maintain the individuality of the molecules,
one could employ low-temperature matrices of rare-gas atoms with molecules
embedded inside the matrix~\cite{AndMos89}. The matrix isolation is a well
established technique in chemical physics. For chemically stable molecules, the
number of guest molecules per host atom (matrix ratio), could be as high as
1/10~\cite{SheridanPrivate}, i.e., with the techniques of matrix isolation, one
could attain the number densities of molecules in the order of $10^{21} \,
\mathrm{cm}^{-3}$.

With the sample number density of $10^{21} \, \mathrm{cm}^{-3}$ and $\mu^\mathrm{CP}$
of BiF~(\ref{Eq:muCPBiFpresent}) derived from the present limit on
eEDM, from Eq.(\ref{Eq:Signal}) we obtain a generated B-field of
$\mathcal{B} \simeq 2 \times 10^{-15} \, \mathrm{Gauss}$. This ultraweak field can be
measured within a month of integration time at the present best sensitivity
limit~\cite{KomKorAll03} of $5 \times 10^{-12} \,
\mathrm{G}/\sqrt{\mathrm{Hz}}$. A relatively small 0.3 cm$^3$-volume sample has
been used in that experiment. The projected theoretical
limit~\cite{KomKorAll03} of this method is $10^{-13} \,
\mathrm{G}/\sqrt{\mathrm{Hz}}$. More optimistic projected sensitivity of $3
\times 10^{-15} \, \mathrm{G}/\sqrt{\mathrm{Hz}}$ is given in
Ref.~\cite{Lam02}. With this projected sensitivity,  we find  that the present
limit on the eEDM~(\ref{Eq:deTl}) may be improved by three orders of
magnitude,
\begin{equation}
d_e(\mathrm{BiF,projected}) \approx 2 \times 10^{-30} \, e \cdot \mathrm{cm}.
\end{equation}

In Ref.~\cite{RavKozDer05}, we have considered a CP-odd magnetization of a
sample of liquid Xe caused by an externally applied electric
field~(\ref{Eq:BetaCPDef}). Here we focused on a similar magnetization due to
 molecular CP-odd magnetic moments and we find
that the BiF experiment has a substantially better sensitivity to eEDM.
This enhancement is due to (i) larger nuclear charge of Bi ($Z=83$) than that
of Xe ($Z=54$) and (ii) much larger E-field applied to heavy atom/ion:
in case of BiF, the internal molecular field is $\sim 4 \times
10^8\, \mathrm{V/cm}$, while in liquid Xe the E-field is
limited by the breakdown strength of $4 \times 10^5\, \mathrm{V/cm}$. This
large difference in the maximum attainable laboratory field and the internal
molecular field~\cite{San67} is exploited  in more conventional searches for
EDMs with
molecules~\cite{HudSauTar02,WilRamLar84,ChoSanHin89,KawBayBic04}.
It is worth emphasizing that the experiment considered here is based on a bulk
magnetization of {\em diamagnetic} molecules, while the conventional searches
for eEDM determine energy splittings in individual {\em paramagnetic}
molecules.

To summarize, we introduced a concept of molecular  CP-violating (T,P-odd)
magnetic moments, $\mu^\mathrm{CP}$. We related these magnetic moments to
eEDM and we estimated
$\mu^\mathrm{CP}$ for a number of diamagnetic polar molecules. We demonstrated
that  $\mu^\mathrm{CP}$ exhibit a strong $Z^5$ dependence on the nuclear charge
of the heavier molecular constituent. Finally, we evaluated a feasibility of
setting a limit on the eEDM by measuring ultra-weak magnetic fields
produced by a polarized sample of diamagnetic molecules. We found that such an
experiment may improve the present limit on the eEDM by several orders
of magnitude.

We would like to thank D. Budker, A. Cronin, M. Romalis, R. Sheridan,  S. Lamoreaux, and L. You
for discussions.  This work was supported in part by  NSF
Grant No.~PHY-0354876, by NIST precision measurement grant, by RFBR Grant
No.~05-02-16914, and by the NSF through a grant to the Institute for
Theoretical Atomic, Molecular, and Optical Physics at Harvard University and
the Smithsonian Astrophysical Observatory.


\end{document}